# Dynamics of Vortex Core Switching in Ferromagnetic Nanodisks

Q.F. Xiao, J. Rudge, and B.C. Choi
*Department of Physics and Astronomy, University of Victoria, Victoria, BC, V8W 3P6, Canada*

Y.K. Hong
*Department of Electrical and Computer Engineering, University of Alabama, Tuscaloosa, AL 35487, USA*

G. Donohoe
*Department of Electrical and Computer Engineering, University of Idaho, Moscow, ID 83844, USA.*

(Dated: July 14, 2006 )

Dynamics of magnetic vortex core switching in nanometer-scale permalloy disk, having a single vortex ground state, was investigated by micromagnetic modeling. When an in-plane magnetic field pulse with an appropriate strength and duration is applied to the vortex structure, additional two vortices, i.e., a circular- and an anti-vortex, are created near the original vortex core. Sequentially, the vortex-antivortex pair annihilates. A spin wave is created at the annihilation point and propagated through the entire element; the relaxed state for the system is the single vortex state with a switched vortex core.

Formation of vortex ground structure in patterned magnetic thin-film elements and its behavior in an external magnetic field has been actively studied during the last few years due to its possible applications in high-density magnetic storage devices [1–11]. More recently the magnetization dynamics of vortex structures in such small elements has received considerable attention [5–11]. The vortex excitation is found to be strongly influenced by element shapes, and several types of eigenmodes have been predicted and experimentally identified [6-8]. In circular disks on micron and submicron length scales, for example, Park and co-workers observed that the low frequency (i.e., sub-GHz) mode, which is due to the gyrotropic motion of the vortex core, shifts to higher frequency modes as the disk diameter decreases [7]. The higher frequency response was attributed to magnetostatic modes of the magnetization outside the vortex core. Another interesting phenomenon occurring in the vortex structures is the switching of the out-of-plane magnetization of the vortex core by an externally applied field. Kikuchi et al. have reported that the spin configuration of the vortex core in a micron-sized permalloy circular disk is very stable, and the magnetization of the core can be switched by a large (~ 2.5 kOe) vertical field in a static case [2]. In a dynamic case, the possible switching of the core magnetization has been inferred by Neudert *et al.*, in which the experimental results acquired using time-resolved Kerr microscopy were compared to micromagnetic simulations [9]. Indications of a switched vortex core magnetization, however, could not be found in experiments, and the mechanism of the core switching has not been further discussed. In this letter, we report a detailed micromagnetic study of the dynamics of vortex core switching for a 500-nm-diameter permalloy (Py) disk of 30 nm thickness using the Object Oriented Micromagnetic Framework (OOMMF) [12]. We found that the vortex core magnetization can be switched by applying an in-plane magnetic field pulse. The switching occurs by means of the formation of additional vortex–antivortex pair and the subsequent annihilation of vortex cores.

The micromagnetic modeling is based on Landau-Lifshitz-Gilbert (LLG) equation [13]. $d\mathbf{M}/dt = -|\gamma|[\mathbf{M}\times\mathbf{H}_{eff}] - \alpha\{\mathbf{M}\times[\mathbf{M}\times\mathbf{H}_{eff}]\}$. Here $\gamma$ is the gyroscopic ratio and $\alpha$ is a phenomenological damping constant. The $\mathbf{H}_{eff}$ is the total effective field acting on the magnetization $\mathbf{M}$, which mainly includes the applied external field $\mathbf{H}_{app}$, the exchange interaction $\mathbf{H}_{ex}$, the magnetic anisotropy field $\mathbf{H}_{ani}$, and the demagnetizing field $\mathbf{H}_d$, i.e., $\mathbf{H}_{eff} = \mathbf{H}_{app} + \mathbf{H}_{ex} + \mathbf{H}_{ani} + \mathbf{H}_d$. The parameters used for

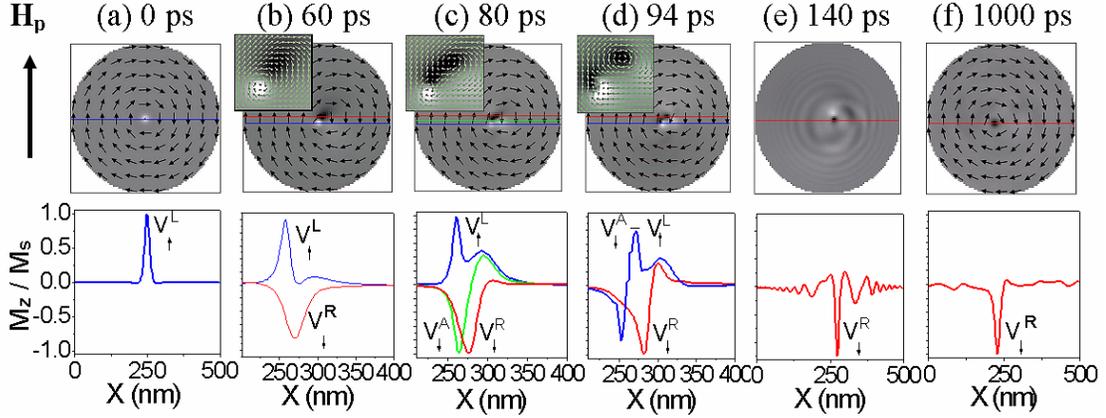

**Fig. 1** (color online) Temporal evolution of $M_z/M_s$ of the Py disk excited by $H_{py}$ = 290 Oe (cut off at 80 ps). The arrows are the in-plane component of the magnetization. Below each image are the Mz/Ms profiles through the centre of each vortex core along the lines in the *x*–direction, respectively. Note that the profiles in (b), (c) and (d) are enlarged in *x*-axis direction.

Py were, saturation magnetization $M_s = 8\times10^5$ A/m, exchange constant $A = 1.3\times10^{-11}$ J/m, and zero magnetic anisotropy constant. In all cases the damping constant of $\alpha = 0.01$ and a cell size of $2\times2\times2$ nm$^3$ were taken. An in-plane magnetic field pulse $\mathbf{H_p}$, with zero rise and fall time, was applied in the positive *y*-direction by varying the strength and duration of the pulse.

Fig. 1 are grayscale images, taken at different times, showing the temporal evolution of the vortex structure in response to the magnetic pulse of 290 Oe. The small arrows in the images represent the in-plane component of the magnetization. Below each image are the $M_z/M_s$ profiles through the centre of each vortex core along the lines in the *x*–direction, respectively. The initial state shown in Fig. 1a contains a single vortex core, with an out-of-plane magnetization in the positive *z*–direction. The vortex structure is identified as left-handed positive ($V_\uparrow^L$) based on its chirality, and positive polarity of the core magnetization [8]. When an in-plane magnetic field pulse is applied, the magnetic system is excited. The initial response of the vortex structure to the external magnetic field is that the spins in the upper half of the disk experience the torque into the plane (i.e., $-z$ direction). This is because the magnetic moments in the upper region are mainly in the positive *x*-direction (Fig.1a). On the contrary, the spins in the bottom half slightly rotate out-of-plane due to the torque in $+z$ direction. Consequently the circular symmetric distribution of the magnetization around the vortex core is distorted, and the spins close to the $V_\uparrow^L$ core leads to the formation of the region (shown as the dark area in Fig.1b), where the spins strongly rotate into the plane in order to avoid the high energetic cost from the anti-parallel in-plane alignments of the nearest spins. This leads to the creation of a new vortex core with the negative polarization ($V_\downarrow^R$), while an antivortex ($V_\downarrow^A$) forms at the boundary of two vortices, $V_\downarrow^R$ and $V_\uparrow^L$ (Fig.1c). With increasing time the vortex-antivortex pair ($V_\uparrow^L$–$V_\downarrow^A$) annihilates (Fig.1d). At this instant the highly concentrated energy in vortex cores is rapidly dissipated into the surrounding area in the form of the spin wave, which is visible in the next two images (Fig.1e and f). As a result of the annihilation, a single vortex core with negative polarity ($V_\downarrow^R$) remains and a vortex core switch is realized (Fig.1f). The switched vortex core, however, still deviates from the central point of

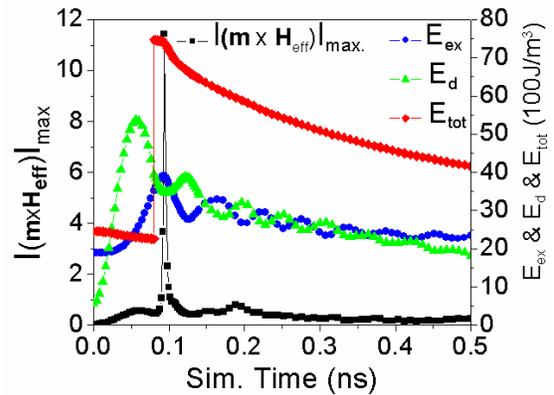

**Fig. 2** (color online) Time dependence of the average energy densities of the exchange energy $E_{ex}$, the demagnetization energy $E_d$, the total energy $E_{tot}$ and the maximum of the nominal torque value $|\mathbf{m} \times \mathbf{h_{eff}}|_{max}$ for the Py disk with a pulse $H_{py}$ = 290 Oe (cut off at 80 ps).

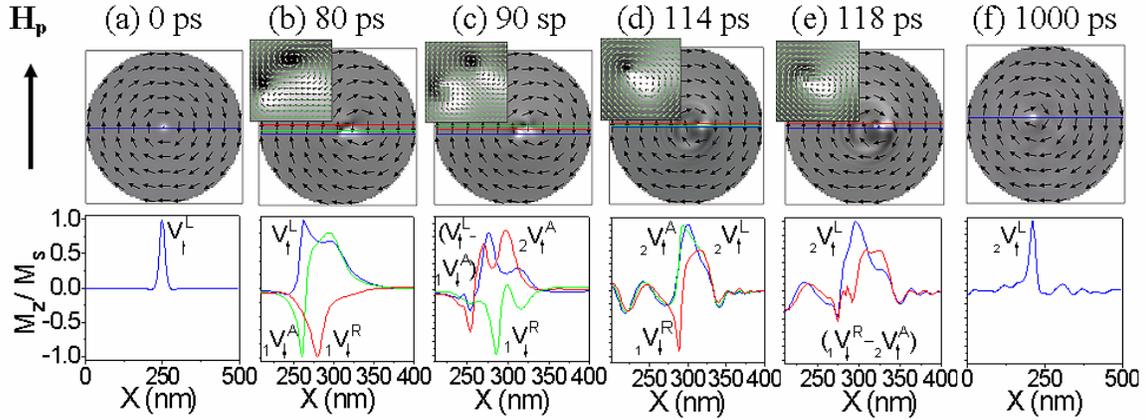

**Fig. 3** (color online) Temporal evolution of $M_z/M_s$ distribution of the Py disk excited by using the pulse of $H_{py} = 350$ Oe (cut off at 80 ps). The arrows are the in-plane component of the magnetization. Below each image are the $M_z/M_s$ profiles through the centre of each vortex core along the lines in the *x*–direction, respectively.

the disk even after a long relaxation, for example as shown at 1 ns in (Fig.1f), implying that the fully relaxed (equilibrium) state has not been reached.

To better understand the mechanism of the vortex core switching, the temporal evolution of the average energy densities, which includes the demagnetization energy $E_d$, the exchange energy $E_{ex}$ and the total energy $E_{tot}$, were inspected. In addition the maximum of the nominal torque value, $|\mathbf{m} \times \mathbf{h}_{eff}|_{max}$ (i.e., $|\mathbf{M} \times \mathbf{H}_{eff}|_{max}/M_s^2$) was also examined. Comparison with Fig. 1 indicates that the change in energies mainly results from the dynamic response of the magnetization configuration around the original vortex core $V_\uparrow^L$. Initially the application of the magnetic field pulse $\mathbf{H}_p$ leads to the displacement of the vortex core ($V_\uparrow^L$) away from the disk center, resulting in the increasing demagnetization energy, while further increase of $E_d$ to a very high energy can be avoided by creating additional vortex cores. It is also noticeable in Fig. 2 that the total energy $E_{tot}$ suddenly drops and there is a sharp peak on the curve of $|\mathbf{m} \times \mathbf{h}_{eff}|_{max}$ when the $V_\uparrow^L$ and $V_\downarrow^A$ pair annihilates. The sudden change of magnetization excites a strong spin wave [12], so that the energy concentrated in the cores, $V_\uparrow^L$ and $V_\downarrow^A$, quickly dissipates into the entire element. Apparently, the energy dissipation through the spin wave propagation speeds up the relaxation process. As analyzed above, this process of core switching depends on the condition of the magnetic field pulse. If the magnetic system is excited by a magnetic pulse with a lower strength than a threshold, additional vortex cores with opposite polarity can not be formed. Whereas, if too high an energy state of the magnetic system is excited by a pulse, it may result in the core switching two or more times.

Fig. 3 shows the evolution of the vortex structure by applying a field pulse of 350 Oe. In response to the pulse, a vortex-antivortex pair ($_1V_\downarrow^R$ and $_1V_\downarrow^A$) is formed while an area with spins oriented in +z-direction (shown as the bright area in Fig. 3b) appears near the $_1V_\downarrow^R$ core. After the pulse is cut off at 80 ps, the first annihilation of the $V_\uparrow^L$–$_1V_\downarrow^A$ pair occurs at 90 ps (Fig.3c), then the bright area separates into the vortex-antivortex ($_2V_\uparrow^L$–$_2V_\uparrow^A$) pair (Fig. 3(d)). Finally, a second annihilation of the $_1V_\downarrow^R$–$_2V_\uparrow^A$ pair occurs at 118 ps, leaving only the vortex core $_2V_\uparrow^L$ with positive polarity. The double annihilation of vortex cores, manifested as the double peaks on $|\mathbf{m} \times \mathbf{h}_{eff}|_{max}$ (Fig. (4)), is equivalent to a non core-switch.

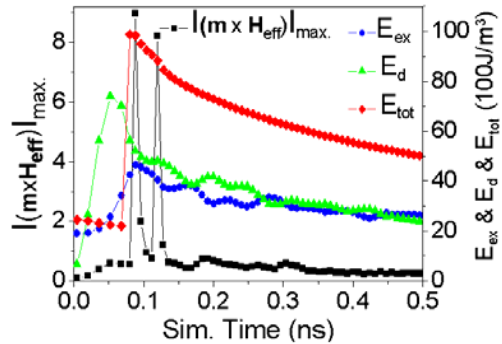

**Fig. 4** (color online) Time dependence of the average energy densities of the exchange energy $E_{ex}$, the demagnetization energy $E_d$, the total energy $E_{tot}$ and the maximum of the nominal torque value $|\mathbf{m} \times \mathbf{h}_{eff}|_{max}$ for the Py disk with an excitation pulse $H_p = 350$ Oe (cut off at 80 ps).

Likewise, keeping the field amplitude at 290 Oe and extending the duration time of the pulse to more than 360 ps, two or more core switches can happen. If the duration of the pulse is too short (< 50 ps), no core switching can occur. This result implies that single or multiple core switches can be controlled by varying the amplitude and duration of the pulse. In more detail, however, the cutoff timing of the pulse also plays a critical role in the core switching due to the oscillatory nature of magnetostatic energy from the gytropic motion of the vortex core [14].

In conclusion, dynamics of vortex core switching in a nanometer-scale permalloy disk has been discussed. We demonstrate that the magnetic vortex core can be reversed via the annihilation of vortex-antivortex pair by applying an in-plane magnetic field pulse with an appropriate strength and duration. From an application point of view, the results suggest a magnetization switching mechanism, which could be applied to magnetic information technologies, from hard drives to spintronic devices.

**Acknowledgments**

This work was supported by the Natural Sciences and Engineering Research Council (NSERC) of Canada and, U.S. Air Force Research Laboratory (AFRL) under Grant No. F29601-04-1-206.

**References**

[1] R. Pulwey, M. Rahm, J. Biberger, and D. Weiss, IEEE Trans. Magn. **37**, 2076 (2001).
[2] N. Kikuchi, S. Okamoto, O. Kitakami, Y. Shimada, S. G. Kim, Y. Otani, and K. Fukamichi, J. Appl. Phys. **90**, 6548 (2001).
[3] T. Okuno, K. Mibu and T. Shinjo, J. Appl. Phys. **95**, 3612 (2004).
[4] A. Thiaville, J. Miguel Garcia, R. Dittrich, J. Miltat, and T. Schrefl, Phys. Rev. B **67**, 94410 (2003).
[5] V. Novosad, M. Grimsditch, K. Yu. Guslienko, P. Vavassori, Y. Otani, and S. D. Bader, Phys. Rev. B **66**, 52407 (2002).
[6] K. Yu. Guslienko, B. A. Ivanov, V. Novosad, Y. Otani, H. Shima, and K. Fukamichi, J. Appl. Phys. **91**, 8037 (2002).
[7] J. P. Park, P. Eames, D. M. Engebretson, J. Berezovsky, and P. A. Crowell, Phys. Rev. B **67**, 20403 (2003).
[8] S.-B. Choe, Y. Acremann, A. Scholl, A. Bauer, A. Doran, J. Stöhr, and H. Padmore, Science **304**, 420 (2004).
[9] A. Neudert, J. McCord, R. Schäfer, and L. Schultz, J. Appl. Phys. **97**, 10E701 (2005).
[10] K. S. Lee, B. W. Kang, Y. S. Yu, and S. K. Kim, Appl. Phys. Lett. **85**, 1568 (2004).
[11] K. S. Lee, S. K. Choi, and S. K. Kim, Appl. Phys. Lett. **87**, 192502 (2005).
[12] M.J. Donahue and D.G. Porter, OOMMF User's Guide, Version **1.0**, NIST (Sept 1999), http://math.nist.gov/oommf/.
[13] L. Landau and E. Lifshitz, Phys. Z. Sowjetunion **8**, (1953) 153; T.L. Gilbert, Phys. Rev. **100**, 1243 (1955).
[14] Q.F. Xiao *et al.*, (*unpublished*)